# Improving Spatial Resolution of First-order Ambisonics Using Sparse MDCT Representation

Denis Likhachov
Computer engineering depratment of
Belarussian State University of
Informatics and Radioelectronics
Minsk, Belarus
likhachov@bsuir.by

Nick Petrovsky
Computer engineering depratment of
Belarussian State University of
Informatics and Radioelectronics
Minsk, Belarus
nick.petrovsky@bsuir.by

Elias Azarov
Computer engineering depratment of
Belarussian State University of
Informatics and Radioelectronics
Minsk, Belarus
azarov@bsuir.by

*Abstract*— The paper presents a method for improving spatial resolution of first-order ambisonic audio. The method is based on time/frequency decomposition of the audio with subsequent extraction of a directed plane wave from each frequency component. The method develops the basic ideas of high angular resolution planewave expansion (HARPEX) and directional audio coding (DirAC) taking advantage of real-valued sparse decomposition. Real-valued frequency components as opposed to complex-valued introduce simpler and more stable direction of arrival estimates, while sparse decomposition introduces an accurate and unified approach to describing sounds of different nature from transient to tonal sounds.

## I. Introduction

First-order ambisonics has become a popular and accessible way of capturing surround sounds. There are now many convenient handheld devices that provide four-channel ambisonic recording that can be played back in virtual reality applications through loudspeakers or headphones. However, quality of rendered audio highly depends on spatial resolution which is rather low for first-order ambisonics. There are parametric methods that produce improved spatial image [1,2] and have proven to be practically effective with remarkable results. The general parametric approach is to perform time/frequency decomposition of the audio and treat frequency components separately representing them as a combination of directional plane waves and/or undirected components. Direction of arrival (DOA) estimates constitute a sharp spatial image that can be encoded into higher-order ambisonics audio or alternative surround sound format – fig. 1.

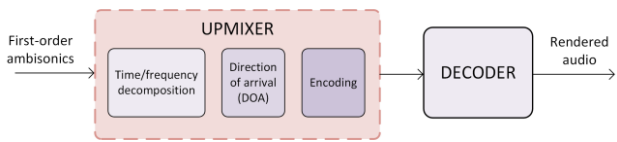

Fig. 1. Rendering ambisonic audio with improved spatial resolution

The present work aims to further develop the approach by implementing some practical ideas concerning time/frequency representation and DOA. Traditionally, fast Fourier transform (FFT) is used for frequency representation of multichannel audio, which is true not only for parametric decoding methods, but also for other applications including noise reduction and dereverberation [3–6]. However, the complex-valued frequency domain is not always advantageous, and in particular we hypothesize that real-valued representations are more suited for determining DAO. As we will show in the paper real-valued Modified discrete cosine transform (MDCT) clearly seems more consistent from a certain point of view. It not only gives a more straightforward interpretation of a directional wave, but also makes the solution more stable because does not have unresolved areas which are specific to HARPEX.

In addition to the transition to the real-valued domain, we develop the idea of using overcomplete bases of different lengths to represent frequencies. The reason for this is possibility of accurate processing of transients (very short components) and tones (long components). When using a linear transform with basis functions of equal length, it is impossible to extract all components with equal accuracy. Usually the choice is made in favor of tones, which leads to blurred transients and significant loss in spatial resolution. In order to perform time/frequency decomposition we use an optimization routine that reduces L1 norm and yields sparse representation. Subjective tests show a good potential of the proposed solution with a drawback of high computational cost.

The paper is organized as follows: In sections II and III we give a HARPEX outline along with proposed solution in order to point out the difference between complex-valued FFT and real-valued MDCT transforms with respect to plane waves extraction. In section IV we present a sparse decomposition routine designed for time/frequency representation. In section V we describe experimental setup and subjective listening results.

## II. HARPEX Outline

HARPEX decomposes 4-channel ambisonic audio into overlapped frames which are transformed into frequency domain using FFT. This yields complex spectral bins of the form $w_r + iw_i$, $x_r + ix_i$, $y_r + iy_i$ and $z_r + iz_i$ for each correspondent channel denoted as $W, X, Y, Z$ respectively. For simplicity, we consider only one bin, assuming that all bins are handled identically. These four complex values are decomposed into two real-valued vectors $[x_1, y_1, z_1]$ and $[x_2, y_2, z_2]$ representing direction of arrival and complex amplitudes $a_1, a_2$ of the sound waves correspondent to these directions. This is given by the following matrix equation:

$$\begin{bmatrix} w_r + iw_i \\ x_r + ix_i \\ y_r + iy_i \\ z_r + iz_i \end{bmatrix} = a_1 \begin{bmatrix} 2^{-\frac{1}{2}} \\ x_1 \\ y_1 \\ z_1 \end{bmatrix} + a_2 \begin{bmatrix} 2^{-\frac{1}{2}} \\ x_2 \\ y_2 \\ z_2 \end{bmatrix} \quad (1)$$

The solution of the system is given in the form:

$$\begin{bmatrix} a_1 \\ a_2 \end{bmatrix} = \begin{bmatrix} m_1 & 0 \\ 0 & m_2 \end{bmatrix} \begin{bmatrix} c_1 & s_1 \\ c_2 & s_2 \end{bmatrix} \begin{bmatrix} 1 \\ i \end{bmatrix} \quad (2)$$

$$c_{1,2} = \sqrt{\frac{2r^2 - pq + p^2 \pm 2r\sqrt{r^2 - pq}}{(q-p)^2 + 4r^2}} \quad (3)$$

$$s_{1,2} = \frac{(q-p)c_{1,2} + p/c_{1,2}}{2r} \quad (4)$$

$$r = -2w_r w_i + x_r x_i + y_r y_i + z_r z_i \quad (5)$$

$$p = -2w_r^2 + x_r^2 + y_r^2 + z_r^2 \quad (6)$$

$$q = -2w_i^2 + x_i^2 + y_i^2 + z_i^2 \quad (7)$$

Some bins fall into uncertainty when $r^2 - pq < 0$, however amount of these bins is rather low.

### III. PROPOSED SOLUTION

In the proposed solution we use MDCT instead of FFT. MDCT gives real-valued bins $w_r, x_r, y_r, z_r$ for each correspondent channel which immediately simplifies the system. The values are decomposed into a real-valued vector $[x_1, y_1, z_1]$ and omnidirectional component with real amplitudes $a_1, a_2$ respectively:

$$\begin{bmatrix} w_r \\ x_r \\ y_r \\ z_r \end{bmatrix} = a_1 \begin{bmatrix} 2^{-\frac{1}{2}} \\ x_1 \\ y_1 \\ z_1 \end{bmatrix} + a_2 \begin{bmatrix} 2^{-\frac{1}{2}} \\ 0 \\ 0 \\ 0 \end{bmatrix} \quad (8)$$

$$a_1 = \sqrt{x_1^2 + y_1^2 + z_1^2} \quad (9)$$

$$a_2 = \sqrt{2} w_r - a_1 \quad (10)$$

The solution in the present form is incomplete, because it loses sign of the direction when $w_r$ is negative. In order to fix that instead of $[x_1, y_1, z_1]$ we can use vector $[\bar{x}_1, \bar{y}_1, \bar{z}_1]$ that considers sign of $w_r$ and correspondent sign-aware amplitudes $\bar{a}_1$ and $\bar{a}_2$:

$$[\bar{x}_1, \bar{y}_1, \bar{z}_1] = \text{sgn}(w_r)[x_r, y_r, z_r] \quad (11)$$

$$\bar{a}_1 = \text{sgn}(w_r) a_1 \quad (12)$$

$$\bar{a}_2 = \sqrt{2} w_r - \bar{a}_1 \quad (13)$$

Thus solution become pretty straightforward and much more simple compared to complex frequency domain. HARPEX extracts two plane waves for each frequency bin and here we have only one plane wave and an omnidirectional component – fig.2.

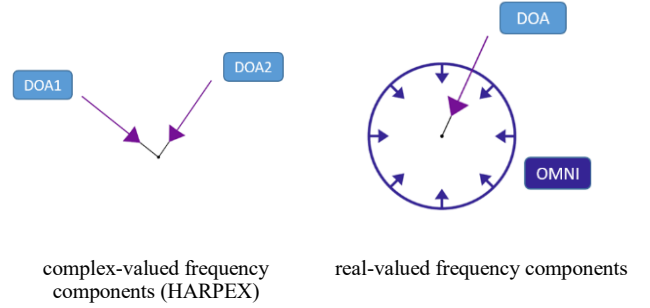

| complex-valued frequency components (HARPEX) | real-valued frequency components |

Fig. 2. Sound wave expansions

The question is, however, what decomposition is better from perceptual perspective. On the one hand, if we split each frequency bin into two waves we can expect a more accurate result. However, we must take into account that in the real-valued case we have twice as many frequency samples. We must also take into account that the spatial separation of two tones of the same frequency is a very difficult task for the auditory system, since they are rather perceived as a single tone. So, theoretically, one would not expect a noticeable difference between the methods.

According to the impression we had during subjective auditions the strongest point of the real-valued decomposition is stability. Because the extracted plane wave is always one-sided, the spatial jitter is reduced. It requires less or even no smoothing of direction vectors or panning weights at decoding stage that yields a sharper audio image. However, it should be taken into account that these impressions are obtained on limited sound material and require more detailed elaboration.

### IV. SPARSE MDCT DECOMPOSITION

#### A. Motivation

A linear time-frequency transform whether FFT or DCT with fixed frame length cannot give an accurate decomposition for both tonal and transient components. These two extremes suffer from blurring in the frequency image. Considering that we perform sparse audio decomposition using a combination of MDCT bases of different lengths. Compared to a fixed linear transform this approach is able to give a good localization for tones and transient impulses – fig. 3.

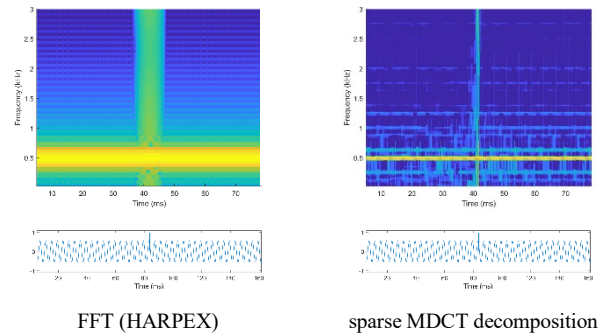

FFT (HARPEX)   sparse MDCT decomposition

## B. Sparse representation combining different time-frequency resolutions

For representing $x$ in real-valued frequency domain we use union of MDCT bases $B$ each with different length:

$$B = \begin{bmatrix} \text{MDCT}_{32} \\ \text{MDCT}_{64} \\ ... \\ \text{MDCT}_{2048} \end{bmatrix}$$

where $\text{MDCT}_N$ stands for a set of MDCT reconstruction functions of length $N$. Representation in each particular basis we will call layer, meaning that the signal is decomposed into separate layers that have different time-frequency characteristics.

Decomposing of the audio comes to the following optimization problem, where $x$ is reconstructed providing the lowest possible L1 norm of representation $X$:

$$\arg\min\{\|X\|_1 \mid x = XB\}$$

where 1-norm $\|X\|_1 = \sum_i |X_i|$ is the sum of absolute values.

There are many known approaches to this problem minimizing L1 norm including

- matching pursuit;
- gradient methods;
- greedy algorithms;
- linear programing.

The most appropriate of all the methods we tried so far is the gradient descent method. The motivation behind this is that on the one hand, the dimensionality of the problem is very large, which does not allow us to use sophisticated calculus such as second derivatives or matrix inversions. On the other hand, a critical part of the solution is ability to suppress distortions caused by aliasing in process of decompositions. Gradient descent method makes it easy to add cost functions in order to achieve aliasing control. The drawback is overall modest performance caused by slow convergence of the solver. However, at the present stage of research, we aim to maximize the quality of the processing while leaving aside performance issues.

## C. Sparse representation solver

We built our experimental gradient descent solver by analogy to neural network training in machine learning. It works by iteratively approaching the solution by stepping towards the direction of the decreasing cost function. The cost function is constituted from three terms: reconstruction loss, 1-norm loss and aliasing loss – fig.4.

Sparsity of the representation is achieved due to 1-norm loss which should be big enough in order to make optimization process faster, but at the same time there is no way to get perfect reconstruction for non-zero 1-norm loss. In order to overcome this we parametrize contribution of the loss with parameter $\alpha$ that starts from a big value that decreases at each iteration, eventually reaching zero.

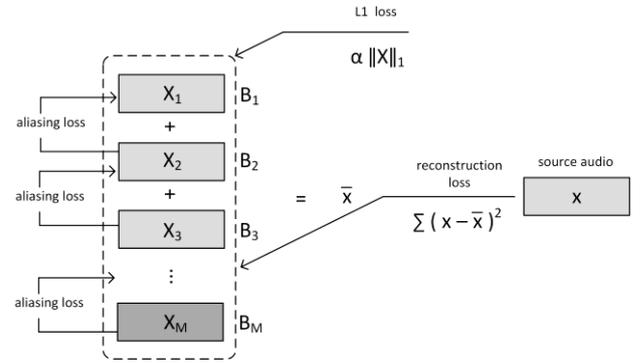

Fig. 4. Designed gradient descent solver

The aliasing cost function is based on the extraction of distortions that appear in layers with longer basis functions. The following principle is pursued: layers with shorter basis functions should not increase the sampling energy of layers with longer basis functions.

## V. EXPERIMENTS

### A. Experimental setup

Audio rendering for listening tests was organized as follows – fig.5. First we upmixed first-order ambisonic audio to seventh-order ambisonics using the following steps: 1) sparse decomposition in MDCT domain; 2) determining direction of arrival; 3) encoding to seventh-order ambisonics. Upmixed audio was decoded into binaural audio using SPARTA binaural decoder [7].

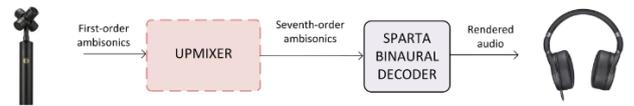

Fig. 5. Experimetal audio rendering with improved spatial reslutions

For sparse decomposition we used five decomposition layers with MDCT sizes 32, 128, 256 1024, 2048 with sine window. In order to speed up optimization routine we used no oversampling in each layer, though it can further improve decomposition quality. The solver performed 2000 iterations producing five decomposition layers containing from shortest transients to long tonal components – Fig.6.

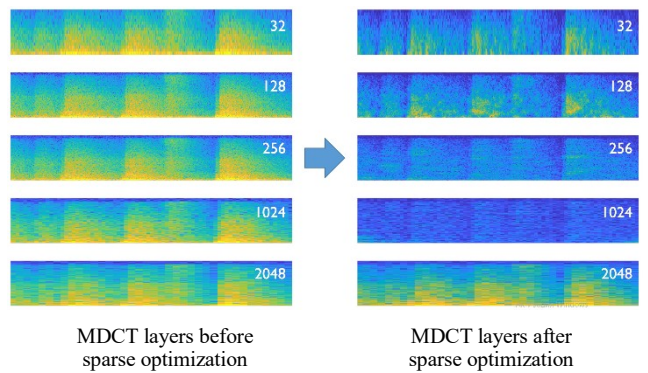

Fig. 6. Sparse decomposition after 2000 iterations

### B. Audio samples

For close analysis and listening tests we used four audio samples recorded with first-order ambisonic devices: 1) foreshore [8]; 2) heavy trucks [8]; 3) elevator [8];

4) orchestral piece [9]. Samples 1–3 were recorded with Rode NT-SF1 microphone and sample 4 with Calrec Soundfield MkIV.

*C. Visual analysis*

Encoded seventh-order ambisonic audio was compared to original visually using EnergyVisualizer plugin from IEM Plug-in Suite created by staff and students of the Institute of Electronic Music and Acoustics [10]. According to sound field visualizations the upmixed audio evidently has a much higher spatial resolution – fig.7.

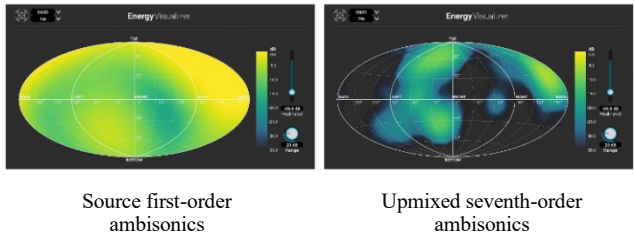

Source first-order ambisonics    Upmixed seventh-order ambisonics

Fig. 7. Sound field visualization of the source and upmixed ambisonics

We found that the spatial visual image is stable and responds to even subtle changes in the position of the sound sources.

*D. Listening tests[1]*

We summarize subjective perception of spatial resolution of the decoded audio samples in Table I. In order to highlight the contribution of each part of the proposed method, we provide estimates for different decoding setups: 1) decoding from first-order ambisonics without upmixing; 2) decoding from seventh-order upmix obtained by using ordinary linear MDCT with length 2048; 3) decoding from seventh-order upmix obtained using sparse MDCT with five decomposition layers and no aliasing cost function; 4) decoding from seventh-order upmix obtained using sparse MDCT with five decomposition layers and aliasing cost function. Listening was carried out in headphones.

TABLE I.     Subjective spatial resolution (MOS)

|  |  | Sample | | | |
| --- | --- | --- | --- | --- | --- |
|  | **Decoding setup** | *1* | *2* | *3* | *4* |
| *1* | First-order | 3.9 | 4.0 | 3.8 | 3.9 |
| *2* | Seventh-order upmix linear MDCT$_{2048}$ | 4.2 | 4.2 | 4.0 | 4.3 |
| *3* | Seventh-order upmix sparse MDCT, no aliasing cost function | 4.5 | 4.4 | 4.3 | 3.5 |
| *4* | Seventh-order upmix sparse MDCT with aliasing cost function | **4.6** | **4.4** | **4.7** | **4.6** |

*E. Discussion*

Though there is an evident improvement in subjective spatial resolution of audio images it should be taken into account that there are limitations of the tests in number of audio samples and listening equipment. Using just headphones may be somewhat inaccurate since decoder uses built-in head-related transfer functions that may not suit well to a particular listener. Regarding audible artifacts, the test is more reliable and indicates that implemented time/frequency decomposition solver provides sparse representation with reasonably good quality. The aliasing cost function makes a significant contribution into perception of some particular sounds as can be clearly seen in sample 4 (orchestral piece) processed with and without aliasing cost function.

A noticeable improvement in spatial resolution is achieved by the sparse representation, as can be seen by comparing modes 2 and 3. In our opinion, this is achieved mainly by separate spatial localization of clear transient sounds, which are quite clearly perceived in the decoded sound image.

## VI. Conclusion

The method presented in the paper improves spatial resolution of the first-order ambisonic audio. The method benefits from real-valued sparse time-frequency decomposition and provides a more sharp spatial image. The practical results indicate applicability of the method for upmixing audio up to seventh-order ambisonics. The disadvantage of the method is a high computational cost, but it is still suitable for applications that do not require real time upmixing.

## Acknowledgment

The authors would like to thank Acoustics Lab at Aalto University and Institute of Electronic Music and Acoustics for making and releasing free spatial audio production tools that have been used extensively in this paper.

---

[1] Some audio samples are given on the presentation slides https://effective-sound.com/downloads/ambisonic_upmix.pptx